\documentclass[12pt]{article}
\begin{document}
\begin{titlepage}
\begin{flushright}
hep-ph/0406207
\end{flushright}
\vspace{1cm}

\begin{centering}
\vspace{.4in} { \Large { \bf Dimensional Reduction of
ten-dimensional\\ \vspace{.3cm} Supersymmetric Gauge Theories\\
\vspace{.3cm} in the ${\cal N}=1$,\ \ $D=4$ Superfield Formalism}}\\
\vspace{1.5cm}

{ \bf P.~Manousselis}$^{1a}$ and { \bf G.~Zoupanos}$^{2b}$\\
\vspace{.2in} $^{1}$Department of Engineering Sciences,\\
University of
Patras, 26110 Patras, Greece. \\
\vspace{.2in} $^{2}$ Physics Department, National Technical
University,
\\ Zografou
Campus, 15780 Athens, Greece.\\

\vspace{1.0in}

{\bf Abstract}\\

\vspace{.1in} A ten-dimensional supersymmetric gauge theory is
written in terms of ${\cal N} =1$, $D=4$ superfields. The theory is
dimensionally reduced over six-dimensional coset spaces. We find
that the resulting four-dimensional theory is either a softly broken
${\cal N}=1$ supersymmetric gauge theory or a non-supersymmetric
gauge theory depending on whether the coset spaces used in the
reduction are non-symmetric or symmetric. In both cases examples
susceptible to yield realistic models are presented.
\end{centering}
\vspace{3.7cm}

\begin{flushleft}
$^{a}$e-mail address: pman@central.ntua.gr.
\\ $^{b}$e-mail address:
George.Zoupanos@cern.ch.
\end{flushleft}
\end{titlepage}
\section{Introduction}
Higher-dimensional ${\cal N}=1$ supersymmetric gauge theories are
known to lead to gauge theories with extended supersymmetries  in
four dimensions. Well known examples are ${\cal N}=1$ supersymmetric
gauge theories in ten and six dimensions leading to ${\cal N}=4$ and
${\cal N}=2$ supersymmetric gauge theories in four dimensions under
trivial reduction \cite{Polchinski:rr}. Four-dimensional gauge
theories with extended supersymmetries are very interesting
frameworks for studying properties of superconformal field theories
and dualities \cite{Aharony:1999ti, Alvarez-Gaume:1996mv}. However,
they are not known to lead to realistic theories describing physics
beyond the Standard Model (SM). On the other hand the SM requires an
understanding of the plethora of its free parameters resulting
mostly from the ad-hoc introduction of the Higgs and Yukawa sectors
in the theory, which could have their origin in a higher-dimensional
theory. Indeed various schemes, with the Coset Space Dimensional
Reduction (CSDR) \cite{Forgacs:1979zs, Kapetanakis:hf, Kuby} being
pioneer have suggested that a unification of the gauge and Higgs
sectors can be achieved in higher dimensions; the four-dimensional
gauge and Higgs fields are simply the surviving components of the
gauge fields of a pure gauge theory defined in higher dimensions. In
the next step of development of the CSDR scheme, fermions were
introduced \cite{Manton:1981es} and then the four-dimensional Yukawa
and gauge interactions of fermions found also a unified description
in the gauge interactions of the higher-dimensional theory. The last
step in this unified description in higher dimensions is to relate
the gauge and fermion fields which have been introduced. This can be
achieved by demanding that the higher-dimensional gauge theory be\
${\cal N}= 1$ supersymmetric; the gauge and fermion fields are
members of the same vector supermultiplet. A very welcome input in
this line of arguments comes from String Theory (for instance the
heterotic string) which fixes the space-time dimension and the gauge
group of the higher-dimensional supersymmetric theory
\cite{Polchinski:rr, Green:mn}. Among the successes of the CSDR
scheme one has to add the possibility of obtaining chiral theories
in four dimensions \cite{Chapline:wy}, as well as softly broken or
non-supersymmetric theories \cite{Manousselis:2001re}.

Motivated by the old work of Marcus, Sagnotti and Siegel (MSS)
\cite{Marcus:1983wb}, a recent article \cite{Arkani-Hamed:2001tb}
which examines the MSS Lagrangian in dimensions between five and
ten and by ref.\cite{Marti:2001iw}, where superspace Lagrangians
with compact extra fifth dimension are considered, we would like
to examine here to which extend our findings in refs.
\cite{Manousselis:2001re} can be described in the superfield
language. We find that indeed this is possible with the exception
of describing the softly supersymmetry breaking gaugino mass term
due to the lack, so far, of a superfield formulation of the
ten-dimensional supergravity-supersymmetric Yang-Mills (SYM)
Lagrangian. For the completeness of presentation we bypass this
difficulty by choosing the torsion of the non-symmetric coset
space such that there is no gaugino mass generated and carrying
out the calculations at the origin of the coset space so that the
form  of the Lagrangian is that of a flat space one. Moreover in
our construction of explicit examples we present the details of a
new calculation leading to the derivation of the potential of the
resulting four-dimensional theory after dimensional reduction.
Finally we show how the four-dimensional GUTs obtained using this
dimensional reduction can potentially become realistic theories.

The present paper is organized as follows. In Section 1 we present
the  $ D=10,\ \ {\cal N}=1 $ Lagrangian in terms of $ D=4, \ \
{\cal N}=1 $ superfields. In Section 2 we perform the CSDR using
four-dimensional ${\cal N} =1$ superfields. In Section 3 using the
superfield formulation we present a  three-generation softly
broken supersymmetric example; it specifically concerns the
reduction of a $G=E_{8}$ SYM theory over $SU(3)/(U(1) \times
U(1))$. In Section 4 we present a  three generation
non-supersymmetric example. This example is obtained by the
reduction of a $G=E_{8}$ SYM theory over $B= (SU(3)/SU(2) \times
U(1)) \times (SU(2)/U(1))$ and the details of the four-dimensional
potential are presented for the first time. Finally Section 5
contains our conclusions.
\section{The $ D=10,\ \ {\cal N}=1 $ Lagrangian in terms of \\ $ D=4,
\  {\cal N}=1 $ superfields}
In ref.~\cite{Marcus:1983wb} the ten-dimensional, ${\cal N}=1$ SYM
Lagrangian was written in terms of four-dimensional ${\cal N} =1$
superfields generalizing earlier formulation of the four-dimensional
${\cal N} =4$ supersymmetric gauge theory in terms of ${\cal N} =1$
superfields \cite{Grisaru:1979wc,Fayet:1978ig}. The Lagrangian
written in superfield formalism  is equivalent to the known
Lagrangian expressed in components \cite{Brink:1976bc}, i.e. it
reduces to the latter when the auxiliary fields are eliminated via
the equations of motion. Then trivial dimensional reduction to four
dimensions, leads to ${\cal N}=4$ supersymmetric gauge theory. In
the present work we perform a generalized dimensional reduction,
namely the CSDR \cite{Forgacs:1979zs}-\cite{Manton:1981es}. We find
that the four-dimensional theory obtained in this way can be a
softly broken ${\cal N }=1$ supersymmetric gauge theory
\cite{Manousselis:2001re} if the reduction is carried over a coset
space $S/R$ which admits an $SU(3)$-structure \cite{Cardoso:2002hd},
a requirement which in turn leads us to the suitable six-dimensional
coset spaces. These are the three well known spaces $G_{2}/SU(3)$,
$Sp(4)/(SU(2) \times U(1))_{non-max}$ and $SU(3)/U(1) \times U(1)$.

It should be stressed that in general in the process of CSDR any
sign of the higher-dimensional supersymmetry may be lost
\cite{Kapetanakis:hf, Chapline:wy, Manousselis:2001re}. Roughly
the reason is the following. We assume a spacetime of the form
$M_{4} \times S/R$ with $M_4$ the four-dimensional Minkowski space
and $S/R$ a coset space. We reduce the theory in $M_4$ by imposing
a generalized $S$-invariance in the extra-dimensional dependence
of the fields defined on $M_{4} \times S/R$, ( the details will be
presented in section~\ref{sec:CSDR}). The effect of this procedure
is that only the $S$-invariant terms in the expansion of the
fields in terms of harmonics of $S/R$ survive. The reason for
obtaining either a non-supersymmetric or a softly broken
supersymmetric theory in four dimensions is due to the fact that
$S$-invariance does not commute with supersymmetry. Either
$S$-invariant bosons and fermions belong to different
representations, as in the former case, or when they belong to the
same representations, as in the latter case, then the interactions
introduced in the process of dimensional reduction break softly
even the remaining ${\cal N }=1$  supersymmetry.

The ${\cal N} = 4$ vector multiplet in ten dimensions decomposes
under ${\cal N} = 1$ as $3$ chiral multiplets $\Phi_a$, $a=1,2,3$
and a vector multiplet $V$. The superspace Lagrangian will be built
out of these fields. These will be ordinary $D=4, {\cal N}=1$
superfields having a dependence on the extra coordinates $y^a$ which
can be considered, from the four-dimensional point of view, just as
parameters. Equivalently the fields $V$ and $\Phi_a$ can be
considered as ordinary fields, i.e. a scalar and a vector on the
transverse space, depending on the parameters $x,\theta,
\bar{\theta}$.


Following  ref.~\cite{Arkani-Hamed:2001tb} we consider the
ten-dimensional ${\cal N}=1$ SYM theory, whose Lagrangian is given
by
\begin{equation}\label{D10N}
{\cal L}=-\frac{1}{4g^{2}}Tr(F_{MN}F^{MN})-\frac{i}{2g^{2}}
Tr(\overline{\lambda}\Gamma^{M}D_{M}\lambda).
\end{equation}
In order to write the theory in terms of $D=4,\ {\cal N}=1$
superfields on a spacetime of the form $M^{4} \times B$ we use the
fact that the six-dimensional transverse space $B$ admits an
almost complex structure which allows us to use complex
coordinates, $z^a,\ a=1,2,3$, at a particular point, which later
will be fixed to be the origin of a coset space $B=S/R$. If $y^1,
y^2, y^3, y^4, y^5, y^6$ are the coordinates of the
six-dimensional transverse space, then we can introduce complex
coordinates by defining
\begin{eqnarray}
\begin{array}{lll}
z^1 =\frac{1}{2}(y^1 + i y^2)& z^2 =\frac{1}{2} (y^3 +i y^4)& z^3
=\frac{1}{2}(y^5 + i y^6)
\end{array}.
\end{eqnarray}
The transverse rotational invariance is the $SO(6)$ rotating the
$y^1, \cdots, y^6$ into each other. The $SU(3)$ subgroup of the
$SO(6) \cong SU(4)$ that rotates  the $z^a$ will be useful in
constructing invariant actions. Similarly to the coordinates we can
use combinations of the real higher-dimensional components of the
gauge field to  construct the complex lowest components of the three
chiral superfields $\Phi_{a}$, as follows
\begin{equation}\label{chi}
\Phi_a |_{\theta=\bar{\theta}=0} = \frac{1}{\sqrt{2}} A_a =
\frac{1}{\sqrt{2}}( A_{4+2a} + i A_{3+2a}).
\end{equation}
We will use the convention $\bar{z}^a = (z_a)^\dagger$ for the
complex coordinates, and $\bar{\Phi}^a = (\Phi_a)^\dagger$ for the
chiral fields.

The ten-dimensional action then takes the form
\cite{Marcus:1983wb,Arkani-Hamed:2001tb}
\begin{eqnarray}\label{Arkani1}
S_{10} &=& \int d^{10}x \Big\{ \int d^2 \theta
 \ Tr \left( \frac{1}{4  g^2}W^\alpha W_\alpha
+ \frac{1}{2  g^2} \epsilon^{abc} \Phi_a(\partial_b \Phi_c -
 \frac{1}{2}[\Phi_b,\Phi_c]) \right) \nonumber \\
& + & \int d^4 \theta\; \frac{1}{ g^2} \ Tr \left( ( \bar{\nabla}^a
+ \bar{\Phi}^a ) e^{-V} (\nabla_a + \Phi_a) e^{V} + \bar{\nabla}^a
e^{-V} \nabla_a e^{V} \right) \Big\}
\\ &+& \nonumber WZW term.
\end{eqnarray}
The choice to express the vector supermultiplet in terms of ${\cal
N }=1$ superfields breaks the original ${\cal R}=SU(4)$ symmetry
of the Lagrangian  down to $SU(3)$. The WZW term vanishes in the
Wess-Zumino gauge. In ref.~\cite{Marcus:1983wb} was proven that
this Lagrangian is equivalent to the on-shell $D=10$, ${\cal N}=1$
SYM given in eq.~(\ref{D10N}).
When eq.~(\ref{Arkani1}) is trivially reduced to four dimensions
the $D=4$, ${\cal N}=4$ Lagrangian is obtained. The six transverse
components of the gauge field become scalars in the adjoint of the
ten-dimensional gauge group $G$. In components it has the form
\begin{eqnarray}\label{n4}
{\cal L}= &-& \frac{1}{4g^{2}} Tr \left( F_{\mu \nu} F^{\mu \nu}+
2D_{\mu}A_{m}D^{\mu}A_{m} - [A_{m},A_{n}]^{2}\right)\nonumber\\
&-&\frac{i}{2g^{2}}Tr(\overline{\lambda}\Gamma^{\mu}D_{\mu}\lambda
+ i\overline{\lambda}\Gamma^{m}[A_{m},\lambda])
\end{eqnarray}
with the potential given by
\begin{equation}
V=-\frac{1}{4g^{2}}Tr\left([A_{m},A_{n}]\right)^{2},
\end{equation}
as can be found by inspection of eq.~(\ref{n4}). In the CSDR
described below we will show how both the gauge symmetry $G$ is
broken to a subgroup $H$ and the ${\cal N}=4$ SUSY is reduced to
softly broken ${\cal N}=1$. Moreover the Higgs fields obtained from
the higher-dimensional components of the gauge field do not belong
in the adjoint representation.

\section{CSDR using four-dimensional ${\cal N} =1$ superfields } \label{sec:CSDR}

Let us proceed by recalling the fundamental ideas of the
dimensional reduction procedure that will be used. Dimensional
reduction is the construction of a lower-dimensional Lagrangian
starting from higher dimensions. In our construction we will  use
the symmetries of the extra dimensions. The CSDR is a dimensional
reduction scheme, in which  the extra dimensions form a coset
space $S/R$ and the symmetry used is the group $S$ of isometries
of $S/R$.

Given the above form of the extra coordinates one can construct a
lower-dimensional Lagrangian by demanding the fields to be
symmetric. This means that one could require the fields to be form
invariant under the action of the group $S$ on extra coordinates,
\begin{equation}\label{eq:frminv}
\delta_{\xi}\Phi^{i}= L_{\xi}\Phi^{i} =0,
\end{equation}
where $L_{\xi}$ is the Lie derivative with respect to the Killing
vectors $\xi$ of the extra dimensional metric. However this
condition appears to be too strong when the higher-dimensional
Lagrangian possesses a symmetry. Then a generalized form of the
symmetry condition~(\ref{eq:frminv}) can be
\begin{equation}
\delta_{\xi}\Phi^{i}=L_{\xi}\Phi^{i} = U_{\xi}\Phi^{i},
\end{equation}
where $U_{\xi}$  is the symmetry of the Lagrangian.

When we apply CSDR in a gauge theory the $U_{\xi}$ is naturally
chosen to be a gauge transformation. Then the original  Lagrangian
becomes independent of the extra coordinates because of the gauge
invariance of the theory.

The original spacetime $(M^{D},g^{MN})$, is assumed to be
compactified to $M^{4}~\times~S/R$, with $S/R$ a coset space. The
original coordinates $\hat{x}^{M}$ become coordinates of $M^{4}
\times S/R$, $\hat{x}^{M}= (x^{\mu}, y^{\alpha})$, where $\alpha$
is a curved index of the coset and  $a$  denotes a flat tangent
space index. The metric is
\begin{equation}
g^{MN}= \left[\begin{array}{cc}\eta^{\mu\nu}&0\\0&-g^{\alpha
\beta}\end{array} \right],
\end{equation}
where $\eta^{\mu\nu}~=~diag(1,-1,-1,-1)$ is the Minkowski
spacetime metric and $g^{\alpha \beta}$ is the coset space metric.

In order to study the geometry of coset spaces it is useful to
divide the generators of $S$, $ Q_{A}$ in two sets the generators
of $R$, $Q_{i}$ $(i=1, \ldots,dimR)$ and the generators of $S/R$,
$ Q_{a}$ $(a=dimR+1 \ldots,dimS)$. Then the commutation relations
for the $S$ generators become
\begin{eqnarray}
\left[ Q_{i},Q_{j} \right] &=& f_{ij}^k Q_{k},\nonumber \\ \left[
Q_{i},Q_{a} \right]&=& f_{ia}^{b}Q_{b},\nonumber\\ \left[
Q_{a},Q_{b} \right]&=& f_{ab}^{i}Q_{i}+f_{ab}^{c}Q_{c} .
\end{eqnarray}
The coset space $S/R$ is called symmetric when $f_{ab}^{c}=0$.
Then the above commutation relations are invariant under the
discrete mapping of generators $Q_{a} \rightarrow -Q_{a}$
\cite{Dolan:2003bj}.

The coordinates $y$ define an element of $S$, $L(y)$, which is a
coset representative. Then the Maurer-Cartan form with values in
the Lie algebra of $S$ is defined by
\begin{equation}
V(y)=L^{-1}(y)dL(y) = e^{A}_{\alpha}Q_{A}dy^{\alpha}
\end{equation}
and obeys the Maurer-Cartan equation,
\begin{equation}\label{eq:MC}
dV+V \wedge V=0.
\end{equation}
From the relation (\ref{eq:MC}), and using standard techniques of
differential geometry, we can develop the geometry of coset spaces
and compute vielbeins, connections, curvature and torsion. For
instance the vielbein and the $R$-connection can be computed at the
origin, $y=0$; the results being $ e^{a}_{\alpha} =
\delta^{a}_{\alpha}$ and $e^{i}_{\alpha} = 0$. More generally in the
CSDR scheme we can perform the necessary calculations at the origin,
$y=0$ of the coset space $S/R$ due to the transitive action of $S$
on $S/R$.

The fermion part of the Lagrangian (\ref{D10N}), when reduced
according to CSDR rules (see e.g. \cite{Kapetanakis:hf}) becomes
\begin{equation}\label{38-6}
L_{Y} =\frac{i}{2}\overline{\psi}\Gamma^{a}\nabla_{a}\psi+
\overline{\psi}V\psi .
\end{equation}
At  $y=0$ one finds that $\nabla_{a}= \phi_{a}$ and therefore the
term $\frac{i}{2}\overline{\psi}\Gamma^{a}\nabla_{a}\psi $ in
eq.~(\ref{38-6}) is exactly the Yukawa term. The last term more
explicitly is given by
\begin{equation}\label{termV}
V=\frac{i}{4}\Gamma^{a}G_{abc}\Sigma^{bc},
\end{equation}
where $\Gamma^{a}$ are the gamma matrices in six dimensions,
$\Sigma^{bc}= 1/2(\Gamma^{a}\Gamma^{b} - \Gamma^{b}\Gamma^{a})$ and
$G_{abc} = (1+k)f_{abc}$, with $k$ a parameter which controls the
torsion and is the term from which the gaugino acquires mass.
Clearly the term (\ref{termV}) can be put equal to zero by the
choice $k=-1$. Then the total action,
\begin{equation}
S=\int d^{4}xd^{d}y\sqrt{-g}\Bigl[-\frac{1}{4}
Tr\left(F_{MN}F_{K\Lambda}\right)g^{MK}g^{N\Lambda}
+\frac{i}{2}\overline{\psi}\Gamma^{M}D_{M}\psi\Bigr] ,
\end{equation}
can be written as \cite{Marcus:1983wb}
\begin{eqnarray}\label{yperpediwn}
S_{10} &=& \int d^{10}x \Big\{ \int d^2 \theta
 \ Tr \left( \frac{1}{4 g^2}W^\alpha W_\alpha
+ \frac{1}{2  g^2} \epsilon^{abc} \Phi_a(\partial_b \Phi_c -
 \frac{1}{2}[\Phi_b,\Phi_c]) \right) \nonumber \\
& + & \int d^4 \theta\; \frac{1}{g^2} \ Tr \left( (
\bar{\partial}^a + \bar{\Phi}^a ) e^{-V} (\partial_a + \Phi_a)
e^{V} + \bar{\partial}^a  e^{-V} \partial_a e^{V} \right) \Big\}.
\end{eqnarray}
The action (\ref{yperpediwn}) is equivalent to the ten-dimensional
SYM on flat space $M_{4} \times T_{y=0}(S/R)$. The action for the
ten-dimensional ${\cal N} =1$ supergravity-SYM theory in terms of
four-dimensional superfields  is not known yet. This would be the
desired form to be used in order to demonstrate the occurrence of
a geometrical mass term for the gaugino in the superfield
language.

Note that in order  to reduce the Lagrangian (\ref{yperpediwn}) over
a coset space $S/R$ we need complex coordinates $z^{a}$,
$z^{\bar{a}}$ at least at the point $y=0$. Also recall that the
subgroup of $SU(4)$ which leaves invariant one supersymmetry is
$SU(3)$. If we assume that the extra dimensions form a compact
space, then the original $SO(6)$ structure group of the frame bundle
should be reducible to $SU(3)$ \cite{Polchinski:rr, Green:mn}. In
order to determine the possible ${\cal N} =1$ reductions we have to
select among the available coset spaces those which admit an
$SU(3)$-structure. Then the original form of the Lagrangian with the
three chiral superfields will be kept in the process of dimensional
reduction. For instance the chiral superfields correspond to $(1,0)$
$1$-forms, the complex conjugate antichiral superfield corresponds
to $(0,1)$ $1$-forms. Then the $(1,0)+(0,1)$-form $\Phi =
\Phi_{a}dz^{a}+\Phi_{\bar{a}}d\bar{z}^{\bar{a}}$ corresponds to the
decomposition $6=3+\bar{3}$ in the embedding  $SO(6) \supset SU(3)$
(See tables 7 and 8). Thus we are led to the known non-symmetric
coset spaces $G_{2}/SU(3), \ Sp(4)/(SU(2) \times U(1))_{non.max.},\
SU(3)/U(1) \times U(1)$. The $SU(3)$-structure is defined through
the invariant tensor $f_{abc}$. An $SU(3)$-structure cannot be
defined for the symmetric coset spaces given that such an invariant
tensor does not exist.

We could have reached to the same conclusions if we have examined
the constraints in component form. In that case we would find that
if the cosets are symmetric then the components of the original
superfields belong to different representations of the gauge group
and therefore no sign of the original supersymmetry would be left in
four dimensions. In the case of non-symmetric cosets we would find a
softly broken supersymmetric theory in four dimensions, i.e.
although the reduction preserves a supersymmetric spectrum, the
interactions that are introduced through the constraints break the
supersymmetry softly.


In the superfield formalism the four-dimensional superfields depend
on extra dimensions. From the four-dimensional point of view the
extra dimensions can be considered as continuous parameters. From
the internal space point of view the vector superfield is just a
scalar function of the coordinates $V(y)$, while the three chiral
superfields constitute a $1$-form  $\Phi_{a}(y)$. This fact makes
easy the formulation of the CSDR scheme using four-dimensional
superfields in the case of non-symmetric coset spaces. Again we
demand that the superfields should be invariant up to a gauge
transformation when $S$ acts on $S/R$. We denote by
$\xi_{A}^{\alpha}$, $A =1,\ldots,dimS$, the Killing vectors which
generate the isometries $S$ of $S/R$ and by $W_{A}$ we denote the
compensating gauge tranformation which corresponds to $\xi_{A}$. We
also define the infinitesimal motion
\begin{equation}
\delta_{A} \equiv L_{\xi_{A}}.
\end{equation}
Then the condition that an infinitesimal motion is compensated by a
gauge transformation takes the following specific form when applied
to the scalar $V$ and to the $1$-form  $\Phi_{a}$ (both belonging to
the adjoint representation of the multidimensional gauge group $G$)
\begin{equation}\label{cct0}
\delta_{A}V = \xi_{A}^{\alpha}\partial_{\alpha}V = [W_{A}, V],
\end{equation}
\begin{eqnarray} \label{cct1}
\delta_{A}\Phi_{\alpha}&=&\xi_{A}^{\beta}\partial_{\beta}\Phi_{\alpha}+\partial_{\alpha}
\xi_{A}^{\beta}\Phi_{\beta}\nonumber\\&=&\partial_{\alpha}W_{A}-[W_{A},\Phi_{\alpha}].
\end{eqnarray}
The $W_{A}$ generally depend on the internal coordinates $y$. The
conditions (\ref{cct0}) and (\ref{cct1}) must be covariant when
$\Phi_{\alpha}(y)$ and $V(y)$ transform under a gauge
tranfomation. We conclude that $W$ transforms as
\begin{equation}\label{Wtr}
\widetilde{W}_{A}=gW_{A}g^{-1}+(\delta_{A}g)g^{-1}.
\end{equation}
The variations  $\delta_{A}$, satisfy the condition
$[\delta_{A},\delta_{B}]=f_{AB}^{\\C}\delta_{C}$. This leads to
another consistency condition on $W$, namely
\begin{equation}\label{Cond1}
\xi_{A}^{\alpha}\partial_{\alpha}W_{B}-\xi_{B}^{\alpha}\partial_{\alpha}
W_{A}+\left[W_{A},W_{B}\right]=f_{AB}^{\ \ C}W_{C}.
\end{equation}
We note that at  the origin, $y=0$ the Killing vectors take the form
\begin{equation}
\xi_{a}^{\alpha} = \delta_{a}^{\alpha}, \ \ \ \ \xi_{i}^{\alpha} =
0,
\end{equation}
which mark out the point $y=0$ as the most suitable point for
performing calculations. From eq.~(\ref{Wtr}) we conclude that
$W_{a}$ can always become zero, while this is not true for $W_{i}$.
Therefore eqs~(\ref{cct0}), (\ref{cct1}) and (\ref{Wtr}) provide us
the freedom to perform our calculations at $y=0$ and to put
$W_{a}=0$. Under these assumptions eq.~(\ref{Cond1}) gives
\begin{eqnarray}\label{Assum}
\partial_{a}W_{b} - \partial_{b}W_{a} &=& f_{abi}W^{i},
\\
\partial_{a}W_{i} &=& 0,\\
\left[ W_{i},W_{j} \right] &=& f_{ij}^{\ \ k}W_{k}.
\end{eqnarray}
The $W_{i}$'s are constant and equal to the generators of the
$R$-subgroup of $G$. The $W_{i}$ will be denoted by  $\Phi_{i}$, and
therefore if $\hat{a} = (1, \ldots, dim \ S)$, we will have
$\Phi_{\hat{a}} = (\Phi_{a}, \ \Phi_{i})$ with $a=(1, \ldots, dim \
S/R)$.

Next we will analyze the constraints (\ref{cct0}) and
(\ref{cct1}). From eq.~(\ref{cct0}) we obtain at  $y=0$
\begin{equation}\label{cct01}
\partial_{a}V =0,\ \ \ \ [W_{i}, V] =0.
\end{equation}
Eq.~(\ref{cct01}) shows that the four-dimensional vector
superfield is independent of the coordinates of the coset space
and belongs to the adjoint representation of the reduced gauge
group $H$ which is the subgroup of $G$ commuting with  $R$, i.e.
$H$ is the centralizer of the image of $R$ in $G$.

In order to reduce the $1$-form $\Phi_{a}$ we use eq.~(\ref{cct1})
and obtain at $y=0$
\begin{equation}\label{cct12}
\partial_{a}\Phi_{b} - \frac{1}{2} f^{ \bar{c} }_{ \ \ a b } \Phi_{ \bar{c} } =
\partial_{b}\Phi_{a} - \frac{1}{2} f^{\bar{c}}_{\
\ b a}\Phi_{\bar{c}}.
\end{equation}
and
\begin{equation}\label{cct11}
[W_{i},\Phi_{a}] = f_{ia}^{\ \ c}\Phi_{c}.
\end{equation}
Eq.~(\ref{cct12}) makes the chiral superfields non-propagating in
the extra dimensions. Eq.~(\ref{cct11}) can be analyzed using
Schur's lemma \cite{Forgacs:1979zs, Kapetanakis:hf} providing  the
rules that determine the chiral superfields  surviving from the
reduction of $\Phi_{a}$. The rules are the following. First we embed
$R$ in $S$ and decompose the adjoint representation of $S$ under
$R$,
\begin{eqnarray} \label{dec1}
S &\supset& R \nonumber \\ adjS &=& adjR+\sum s_{i}.
\end{eqnarray}
Then we embed $R$ in  $G$ and decompose the adjoint of $G$ under
$R \times H $,
\begin{eqnarray} \label{dec2}
G &\supset& R_{G} \times H \nonumber \\
 adjG &=&(adjR,1)+(1,adjH)\nonumber\\&&+\sum(r_{i},h_{i}).
\end{eqnarray}
The rule is that when $$s_{i}=r_{i},$$ i.e. when there exist two
identical representations of $R$ in the decompositions
(\ref{dec1}) and (\ref{dec2}), there is one $h_{i}$ multiplet of
chiral superfields surviving in four dimensions. If  $Q_{ax}$ are
the generators of  $G$ belonging to the  $\sum(r_{i},h_{i})$ part
of the decomposition, i.e.  $a$ is an $r_{i}$ index and $x$ is an
$h_{i}$ index, then the unconstrained superfields are $B^{x}$ with
\begin{equation}
\Phi_{a} =B^{x} Q_{ax}.
\end{equation}




Next we reduce the action (\ref{yperpediwn}). The superpotential
${\cal W}$ is
\begin{equation}\label{SP}
{\cal W} = \epsilon^{abc}\Phi_{a}(\partial_{b}\Phi_{c} -
\frac{1}{2}[\Phi_{b},\Phi_{c}])
\end{equation}
and can be written in a more transparent form in order to use the
CSDR constraints, as
\begin{equation}\label{SP1}
{\cal W} = \frac{1}{2}\epsilon^{abc}\Phi_{a}(\partial_{b}\Phi_{c}
-
\partial_{c}\Phi_{b}) -\frac{1}{2}
\epsilon^{abc}\Phi_{a}[\Phi_{b},\Phi_{c}].
\end{equation}
Then using the constraints of CSDR we obtain
\begin{equation}
\partial_{b}\Phi_{c} = \partial_{b}W_{c}+\frac{1}{2}f_{\
bcd}\bar{\Phi}^{d},
\end{equation}
and antisymmetrizing the indices  $bc$ we obtain
\begin{equation}\label{AS}
\partial_{b}\Phi_{c}-\partial_{c}\Phi_{b} = \partial_{c}W_{b}-\partial_{b}W_{c}+f_{\
bcd}\bar{\Phi}^{d}.
\end{equation}
Next due to the constraint  (\ref{Assum}) we obtain,
\begin{equation}\label{CSTR}
\partial_{c}W_{b}-\partial_{b}W_{b} = f_{\ bci}\bar{\Phi}^{i}
\end{equation}
and the superpotential (\ref{SP1}) takes the form
\begin{equation}\label{SuP}
{\cal W} = \frac{1}{2} Tr \left( \epsilon^{abc}\Phi_{a}(f_{\
bc\hat{d}}\bar{\Phi}^{\hat{d}} - [\Phi_{b}, \Phi_{c}]) \right).
\end{equation}
Setting $(M)^{a}_{\hat{b}} \equiv \epsilon^{abc}f_{\hat{b}b c} $
we see that a non-holomorphic term
\begin{equation}\label{dager}
(M)^{a}_{b} \Phi_{a} \bar{\Phi}^{b}
\end{equation}
has occurred in ${\cal W}$, in the process of dimensional reduction,
which breaks supersymmetry and provides the four-dimensional
Lagrangian with scalar mass soft terms. On the other hand, we
identify as superpotential $W$ of the four-dimensional theory the
holomorphic part of the original superpotential, ${\cal W}$
\begin{equation}\label{holsup}
W = -  \frac{1}{2} Tr \left( \epsilon^{abc}\Phi_{a} [\Phi_{b},
\Phi_{c}] \right).
\end{equation}

To determine the contribution of the term (\ref{dager}) we should
compute $ M \bar{\Phi} \Phi|_{\theta^{2}} $. Using the form of
 $\Phi$ in components,
\begin{equation}
\Phi(x,y) = \phi(x,y) + \theta^{2}F(x,y)+ \ldots
\end{equation}
and
\begin{equation}
\bar{\Phi} = \phi^{\ast}(x,y) + \bar{\theta}^{2} F^{\ast}(x,y)+
\ldots ,
\end{equation}
we conclude that
$$  M  \bar{\Phi} \Phi|_{\theta^{2}} = M F \phi^{\ast}.$$
From the equations of motion for  $F$ we obtain
\begin{equation}\label{aster}
F^{\ast} = - \left( \frac{\partial W}{\partial \phi} \phi + M
\phi^{\ast} \right).
\end{equation}
Then substituting $F^{\ast}$, as given in eq.~(\ref{aster}), in the
Lagrangian expressed in components and specifically in the terms
$$ \frac{1}{2} F^{\ast} F + \frac{1}{2} \frac{\partial
\overline{W}}{\partial \phi^{\ast}}F^{\ast} + M F^{\ast} \phi$$  we
obtain finally the familiar supersymmetric $F$-terms plus terms
which break supersymmetry softly, i.e.
\begin{equation}\label{plus}
\left|\frac{\partial W}{\partial \phi} \right|^{2} + M
\frac{\partial W}{\partial \phi} \phi + M^{2} \phi^{2}.
\end{equation}

Next we examine the interaction term
\begin{equation}\label{kovacs1}
 (\bar{\partial}^a + \bar{\Phi}^a ) e^{-V} (
\partial_a + \Phi_a)  e^{V} +  \bar{\partial}^a  e^{-V}
\partial_a e^{V}.
\end{equation}
We shall show how the usual kinetic term of ${\cal N}=1$ theories
\begin{equation}\label{kovacs3}
Tr\left(\bar{\Phi}e^{-V} \Phi \right)
\end{equation}
is obtained.
In order to reduce the kinetic term we consider the relevant part of
eq.~(\ref{yperpediwn})
\begin{equation}\label{KTerm}
 Tr \left( \bar{\Phi}^a  e^{-V} \Phi_a e^{V} \right)
\end{equation}
and write the components in the Wess-Zumino gauge
\begin{equation}
\bar{\Phi}^a(1-V+1/2V^{2})\Phi_a(1+V+1/2V^{2}),
\end{equation}
where both the  vector and chiral superfields belong to the
adjoint representation of $G$
\begin{equation}
\Phi_{a} = \Phi_{a}^{\alpha} T^{\alpha},\ \ V= V^{\alpha}
T^{\alpha}.
\end{equation}
Then, taking into account the commutator $[T^{\alpha}, T^{\beta}] =
g^{\alpha \beta \gamma}T^{\gamma}$, we find
\begin{equation}
\bar{\Phi}^a \Phi_a - g_{\alpha \beta \gamma} \bar{\Phi}^{a \alpha}
V^{\beta} \Phi_{a}^{\gamma} + \frac{1}{2}g_{\alpha
\beta}^{\epsilon}g_{\epsilon \gamma \delta} \bar{\Phi}^{a \alpha}
V^{\beta} V^{\gamma} \Phi_{a}^{\delta} .
\end{equation}
When the representations of $G$ are reduced in those of $H$, the
structure constants $g_{\alpha \beta \gamma}$ of $G$ are reduced
accordingly to the structure constants of $H$.
The kinetic term thus
obtained is the familiar one,
\begin{equation}
Tr \bar{\Phi}(e^{-V})\Phi.
\end{equation}
Let us see this point in some more detail. Initially we had chiral
superfields $\Phi_{a}^{\alpha}$ with $a=1,2,3$ and $\alpha = 1
\ldots dim\ G$ an index in the adjoint representation of the
higher-dimensional gauge group $G$. Through the reduction procedure
we obtain $\Phi_{a}^{\alpha} \rightarrow \Phi_{\hat{a}}^{x}$ with
$\hat{a} $ counting the number of the surviving generations; it
depends on the non-symmetric coset over which we reduce the theory.
Specifically $\hat{a}=1$ if we reduce the theory over $G_{2}/SU(3)$,
$\hat{a}=1,2 $ if we reduce the theory over $Sp(4)/(SU(2) \times
U(1))_{non.max}$ and $\hat{a}= 1, 2, 3 $ if we reduce the theory
over $SU(3)/U(1) \times U(1)$ \cite{Manousselis:2001re}. The index
$x$  denotes the $H$-representation in which the surviving chiral
superfields belong. We will further denote the adjoint of the
reduced gauge group $H$ with an index $i$. Therefore the reduction
can be represented as
$$ \Phi^{a \alpha}(f^{\gamma})_{\alpha}^{\beta} V^{\gamma} \Phi_{a
\beta} \rightarrow \Phi^{\hat{a} x}(f^{i})_{x}^{y} V^{i}
\Phi_{\hat{a} y}.$$

A final remark concerning the dimensional reduction described in the
present section is the following. According to the rules of the
reduction obtained from the constraints we have to embed $R$ in $G$.
In turn the embedding determines the four-dimensional gauge group
$H$ as being the centralizer of  $R$ in $G$, i.e. $H=C_{G}(R)$. A
second role of this embedding, or more accurately of identifying $R$
with a subgroup of $G$, is that it provides a non-trivial background
configuration which could result in obtaining chiral fermions in
four dimensions. The index of the Dirac operator is a topological
invariant depending only on the coset space and the background gauge
fields \cite{Kapetanakis:hf, Dolan:2003bj}.

Usually one considers a simple background configuration and the
chirality index is related with the Euler characteristic of a given
coset space \cite{Lust:1986ix, Dolan:2003bj, Kapetanakis:hf} (see
however \cite{Govindarajan:1986iz}). The point we would like to
stress here is that, in principle, one could consider background
configurations with non-trivial winding number. This offers the
possibility to obtain more fermion families in four dimensions, a
fact particularly useful when one considers reduction over coset
spaces which are multiply connected.

Multiply connected spaces in the present framework can result by
moding out a freely acting discrete group from a coset space and
can be used to break the gauge group using the so called Wilson
flux or Hosotani mechanism \cite{Hosotani:1983xw, Zoupanos:1987wj,
Kapetanakis:hf}. Therefore, what we should do in order to use the
possibility offered by considering a background configuration with
winding number $n$ is to perform a large gauge transformation,
which will take us from one vacuum to another with different
winding number, followed by a small gauge transformation which
will compensate the change of field in higher dimensions as was
described in the present section. The first gauge transformation
will provide the necessary chiral fermion multiplicity and the
latter the geometrical breaking of the gauge group $G$ down to $H$
in four dimensions.

\section{A three generation softly broken supesymmetric example: Reduction
of the $G=E_{8}$ over $SU(3)/(U(1) \times U(1))$.} In this section,
making use of the reduction in the superfield formalism developed in
section 3, we present explicitly the reduction of a $G=E_{8}$ SYM
theory over the coset $SU(3)/(U(1) \times U(1))$. This reduction has
been  carried out in components in ref.~\cite{Manousselis:2001re}.
The decomposition to be used is
$$ E_{8}  \supset U(1)
\times U(1) \times E_{6} $$ The $248$ representation of $E_{8}$ is
decomposed
 under $U(1) \times U(1) \times E_{6}$ as
\begin{eqnarray}\label{78}
 248 &=& 1_{(0,0)}+1_{(0,0)}+1_{(3,\frac{1}{2})}+1_{(-3,\frac{1}{2})}+\nonumber\\
& &1_{(0,-1)}+1_{(0,1)}+1_{(-3,-\frac{1}{2})}+1_{(3,-\frac{1}{2})}+\nonumber\\
&&78_{(0,0)}+27_{(3,\frac{1}{2})}+27_{(-3,\frac{1}{2})}+27_{(0,-1)}+\nonumber\\
&&\overline{27}_{(-3,-\frac{1}{2})}+\overline{27}_{(3,-\frac{1}{2})}
+\overline{27}_{(0,1)}.
\end{eqnarray}
In the present case $R$ is chosen to be identified with the $U(1)
\times U(1)$ of the decomposition (\ref{78}). Therefore the
resulting four-dimensional gauge group, according to the rule stated
in eq.~(\ref{cct01}), is
$$ H=C_{E_{8}}(U(1) \times U(1)) =
 U(1) \times U(1) \times E_{6}, $$
i.e. we find that the surviving fields in four dimensions are
three ${\cal N}=1$ vector multiplets $V^{\alpha},V_{(1)},V_{(2)}$,
(where $\alpha$ is an $E_{6}$, $78$ index and the other two refer
to the two $U(1)'s$) containing the gauge fields of $U(1) \times
U(1) \times E_{6}$, i.e. the original vector superfield $V^{A}$
(where $A$ a $248$ index of $E_{8}$) gives
$V^{\alpha},V_{(1)},V_{(2)}$.

The $R=U(1) \times U(1)$ content of $SU(3)/U(1) \times U(1)$ vector
is according to table 8,
$$(3,\frac{1}{2})+(-3,\frac{1}{2})
+(0,-1)+(-3,-\frac{1}{2})+(3,-\frac{1}{2})+(0,1).$$
Next we apply the solutions given in
eqs.~(\ref{dec1}),(\ref{dec2}) on the surviving chiral superfields
of the constraint (\ref{cct11}) and we find the following answer.
The matter content of the four-dimensional theory consists of
three ${\cal N}=1$ chiral multiplets ($A^{i}$, $B^{i}$, $C^{i}$)
with $i$ an $E_{6}$, $27$ index and three ${\cal N}=1$ chiral
multiplets ($A$, $B$, $C$) which are $E_{6}$ singlets and carry
$U(1) \times U(1)$ charges i.e. the original chiral superfields
$\Phi^{A}_{a}$ (where $A$ is a $248$ index of $E_{8}$ as before
and $a=1, 2, 3$) give ($A^{i}$, $B^{i}$, $C^{i}$ $A$, $B$, $C$).

To understand the above result in detail we examine further the
decomposition of the adjoint of the specific $S=SU(3)$ under $R=U(1)
\times U(1)$, i.e.
$$SU(3) \supset U(1) \times U(1) $$
\begin{eqnarray}\label{dec79}
 8 = (0,0)+(0,0)+(3,\frac{1}{2})+(-3,\frac{1}{2})
+(0,-1)+\nonumber\\(-3,-\frac{1}{2})+(3,-\frac{1}{2})+(0,1).
\end{eqnarray}
Then according to the decomposition (\ref{dec79}) the generators
of $SU(3)$ can be grouped as
\begin{equation}\label{80}
Q_{SU(3)} = \{Q_{0},Q'_{0},Q_{1},Q_{2},Q_{3},Q^{1},Q^{2},Q^{3} \}.
\end{equation}
( The non trivial commutator relations of $SU(3)$ generators
(\ref{80}) are given in table 1). The decomposition (\ref{80})
suggests to introduce the following  notation for the
four-dimensional constrained chiral superfields
\begin{equation}\label{81}
( \Phi_{1}, \Phi^{1}, \Phi_{2}, \Phi^{2}, \Phi_{3}, \Phi^{3}),
\end{equation}
with $\Phi^{a} \equiv \bar{\Phi}^{a} = (\Phi_{a})^{\dagger}$,
$a=1,2,3$.

Next we examine the commutation relations of $E_{8}$ under the
decomposition (\ref{78}). Under this decomposition the generators
of $E_{8}$ can be grouped as
\begin{eqnarray}\label{84}
Q_{E_{8}}=\{Q_{0},Q'_{0},Q_{1},Q_{2},Q_{3},Q^{1},Q^{2},Q^{3},Q^{\alpha},\nonumber\\
Q_{1i},Q_{2i},Q_{3i},Q^{1i},Q^{2i},Q^{3i} \},
\end{eqnarray}
where, $ \alpha=1,\ldots,78 $ and $ i=1,\ldots,27 $. The non-trivial
commutation relations of the $E_{8}$ generators (\ref{84}) are given
in tables 2 and 3.  Now the constraints (\ref{cct11}) of the
reduction scheme on the superfields given in (\ref{81}) can be
expressed as
\begin{eqnarray}\label{85}
\left[\Phi_{1}, \Lambda Q_{0}\right]=\sqrt{3}\Phi_{1}&,&
\left[\Phi_{1},\Lambda' Q_{0}'\right]=\Phi_{1}, \nonumber \\
\left[\Phi_{2},\Lambda Q_{0}\right]=-\sqrt{3}\Phi_{2}&,&
\left[\Phi_{2},\Lambda' Q_{0}'\right]=\Phi_{2}, \nonumber \\
\left[\Phi_{3},\Lambda Q_{0}\right]=0&,& \left[\Phi_{3}, \Lambda'
Q_{0}'\right]=-2\Phi_{3}.
\end{eqnarray}
Then the solutions of the constraints (\ref{85}) in terms of the
genuine four-dimensional chiral superfields and of the $E_{8}$
generators (\ref{84}), corresponding to the embedding (\ref{78}) of
$R=U(1) \times U(1)$ in the $E_{8}$, are
\begin{eqnarray}
\Phi_{1} &=& R_{1} A^{i} Q_{1i}+R_{1} A Q_{1}, \nonumber\\
\Phi_{2} &=& R_{2} B^{i} Q_{2i}+ R_{2} B Q_{2}, \nonumber\\
\Phi_{3} &=& R_{3} C^{i} Q_{3i}+ R_{3} C Q_{3},
\end{eqnarray}
with $\Lambda=\Lambda'=\frac{1}{\sqrt{10}}$. Moreover the
unconstrained chiral superfields transform under $E_{6} \times U(1)
\times U(1)$ as
\begin{eqnarray}
A_{i} \sim 27_{(3,\frac{1}{2})}&,&A \sim
1_{(3,\frac{1}{2})},\nonumber\\ B_{i} \sim
27_{(-3,\frac{1}{2})}&,&B \sim 1_{(-3,\frac{1}{2})},\nonumber\\
C_{i} \sim 27_{(0,-1)}&,&C \sim 1_{(0,-1)}.
\end{eqnarray}

In terms of the unconstrained superfields the superpotential
(\ref{holsup}) takes the form
\begin{equation}\label{sup61}
W(A^{i},B^{j},C^{k},A,B,C) =\sqrt{40}d_{ijk}A^{i}B^{j}C^{k} +
\sqrt{40}ABC.
\end{equation}
In addition the trilinear and mass terms which break supersymmetry
softly can be read from the two last terms in eq.~(\ref{plus}) and
are given by
\begin{eqnarray}\label{scalarSSB} \lefteqn{{\cal L}_{scalarSSB}=
\biggl( \frac{4R_{1}^{2}}{R_{2}^{2}R_{3}^{2}}-\frac{8}{R_{1}^{2}}
\biggr)\alpha^{i}\alpha_{i} +\biggl(
\frac{4R_{1}^{2}}{R_{2}^{2}R_{3}^{2}}-\frac{8}{R_{1}^{2}}
\biggr)\overline{\alpha}\alpha} \nonumber\\ & &
+\biggl(\frac{4R_{2}^{2}}{R_{1}^{2}R_{3}^{2}}-\frac{8}{R_{2}^{2}}\biggr)
\beta^{i}\beta_{i}
+\biggl(\frac{4R_{2}^{2}}{R_{1}^{2}R_{3}^{2}}-\frac{8}{R_{2}^{2}}\biggr)
\overline{\beta}\beta +\biggl(\frac{4R_{3}^{2}}{R_{1}^{2}R_{2}^{2}}
-\frac{8}{R_{3}^{2}}\biggr)\gamma^{i}\gamma_{i}
+\biggl(\frac{4R_{3}^{2}}{R_{1}^{2}R_{2}^{2}}
-\frac{8}{R_{3}^{2}}\biggr)\overline{\gamma}\gamma \nonumber\\ & &
+\biggl[\sqrt{2}80\biggl(\frac{R_{1}}{R_{2}R_{3}}+\frac{R_{2}}{R_{1}
R_{3}}+\frac{R_{3}}{R_{2}R_{1}}\biggr)d_{ijk}\alpha^{i}\beta^{j}\gamma^{k}
\nonumber \\ &
&+\sqrt{2}80\biggl(\frac{R_{1}}{R_{2}R_{3}}+\frac{R_{2}}{R_{1}
R_{3}}+\frac{R_{3}}{R_{2}R_{1}}\biggr)\alpha\beta\gamma+ h.c\biggr],
\end{eqnarray}
where $\alpha^{i}$ etc are the scalar components of the
corresponding superfields. The superpotential (\ref{sup61}), the
$D$-terms and the scalar soft supersymmetry terms given in
eq.~(\ref{scalarSSB}) provide the four-dimensional theory with a
potential as follows
\begin{eqnarray}\label{88}
V(\alpha^{i},\alpha,\beta^{i},\beta,\gamma^{i},\gamma)= const. +
\biggl( \frac{4R_{1}^{2}}{R_{2}^{2}R_{3}^{2}}-\frac{8}{R_{1}^{2}}
\biggr)\alpha^{i}\alpha_{i} +\biggl(
\frac{4R_{1}^{2}}{R_{2}^{2}R_{3}^{2}}-\frac{8}{R_{1}^{2}}
\biggr)\overline{\alpha}\alpha \nonumber \\
+\biggl(\frac{4R_{2}^{2}}{R_{1}^{2}R_{3}^{2}}-\frac{8}{R_{2}^{2}}\biggr)
\beta^{i}\beta_{i}
+\biggl(\frac{4R_{2}^{2}}{R_{1}^{2}R_{3}^{2}}-\frac{8}{R_{2}^{2}}\biggr)
\overline{\beta}\beta \nonumber \\
+\biggl(\frac{4R_{3}^{2}}{R_{1}^{2}R_{2}^{2}}
-\frac{8}{R_{3}^{2}}\biggr)\gamma^{i}\gamma_{i}
+\biggl(\frac{4R_{3}^{2}}{R_{1}^{2}R_{2}^{2}}
-\frac{8}{R_{3}^{2}}\biggr)\overline{\gamma}\gamma \nonumber\\
+\biggl[\sqrt{2}80\biggl(\frac{R_{1}}{R_{2}R_{3}}+\frac{R_{2}}{R_{1}
R_{3}}+\frac{R_{3}}{R_{2}R_{1}}\biggr)d_{ijk}\alpha^{i}\beta^{j}\gamma^{k}\nonumber\\
+\sqrt{2}80\biggl(\frac{R_{1}}{R_{2}R_{3}}+\frac{R_{2}}{R_{1}
R_{3}}+\frac{R_{3}}{R_{2}R_{1}}\biggr)\alpha\beta\gamma+
h.c\biggr]\nonumber\\
+\frac{1}{6}\biggl(\alpha^{i}(G^{\alpha})_{i}^{j}\alpha_{j}
+\beta^{i}(G^{\alpha})_{i}^{j}\beta_{j}
+\gamma^{i}(G^{\alpha})_{i}^{j}\gamma_{j}\biggr)^{2}\nonumber\\
+\frac{10}{6}\biggl(\alpha^{i}(3\delta_{i}^{j})\alpha_{j} +
\overline{\alpha}(3)\alpha + \beta^{i}(-3\delta_{i}^{j})\beta_{j}
+ \overline{\beta}(-3)\beta \biggr)^{2}\nonumber \\
+\frac{40}{6}\biggl(\alpha^{i}(\frac{1}{2}\delta_{i}^{j})\alpha_{j}
+ \overline{\alpha}(\frac{1}{2})\alpha +
\beta^{i}(\frac{1}{2}\delta^{j}_{i})\beta_{j} +
\overline{\beta}(\frac{1}{2})\beta +
\gamma^{i}(-1\delta_{i}^{j})\gamma_{j} +
\overline{\gamma}(-1)\gamma \biggr)^{2}\nonumber \\
+40\alpha^{i}\beta^{j}d_{ijk}d^{klm}\alpha_{l}\beta_{m}
+40\beta^{i}\gamma^{j}d_{ijk}d^{klm}\beta_{l}\gamma_{m}
+40\alpha^{i}\gamma^{j}d_{ijk}d^{klm}\alpha_{l}\gamma_{m}\nonumber\\
+40(\overline{\alpha}\overline{\beta})(\alpha\beta) +
40(\overline{\beta}\overline{\gamma})(\beta\gamma) +
40(\overline{\gamma}\overline{\alpha})(\gamma\alpha).
\end{eqnarray}
Note that the potential (\ref{88}) belongs to the case that $S$ has
an image in $G$. Here $S=SU(3)$ has an image in $G=E_{8}$ and
therefore we conclude that the minimum of the potential is zero and
the four-dimensional gauge group is $E_{6}$ \cite{Kapetanakis:hf}.
To break this gauge group further we should employ the Wilson flux
mechanism and study the potential (\ref{88}). Given that the Euler
number of the coset space $SU(3)/ U(1) \times U(1)$ is $6$,
therefore leading to three families in four-dimensions, the use of
Wilson flux mechanism requires the introduction of a background
configuration with non-trivial winding number appropriate to keep
unchanged the number of families. We plan to discuss the details and
the phenomenological consequences of such construction in a
forthcoming publication. \\

\begin{tabular}{|l|l|l|}\hline
\multicolumn{3}{|c|}{Table 1}\\ \multicolumn{3}{|c|}{Non-trivial
commutation relations of $SU(3)$ according to} \\
\multicolumn{3}{|c|}{the decomposition given in eq.(\ref{80})}\\
\hline $\left[Q_{1},Q_{0}\right]=\sqrt{3}Q_{1}$ &
$\left[Q_{1},Q'_{0}\right]=Q_{1}$ &
$\left[Q_{2},Q_{0}\right]=-\sqrt{3}Q_{2}$\\
$\left[Q_{2},Q'_{0}\right]=Q_{2}$ & $\left[Q_{3},Q_{0}\right]=0$ &
$\left[Q_{3},Q'_{0}\right]=-2Q_{3}$ \\
$\left[Q_{1},Q^{1}\right]=-\sqrt{3}Q_{0}-Q'_{0}$ &
$\left[Q_{2},Q^{2}\right]=\sqrt{3}Q_{0}-Q'_{0}$ &
$\left[Q_{3},Q^{3}\right]=2Q'_{0}$ \\
$\left[Q_{1},Q_{2}\right]=\sqrt{2}Q^{3}$ &
$\left[Q_{2},Q_{3}\right]=\sqrt{2}Q^{1}$ &
$\left[Q_{3},Q_{1}\right]=\sqrt{2}Q^{2}$\\ \hline
\end{tabular}\\ \vspace{.2in}

The normalization of generators in the above table is
$$Tr(Q_{0}Q_{0})=Tr(Q'_{0}Q'_{0})=Tr(Q_{1}Q^{1})=Tr(Q_{2}Q^{2})=Tr(Q_{3}Q^{3})=2$$

\begin{tabular}{|l|l|l|}\hline
\multicolumn{3}{|c|}{Table 2}\\  \multicolumn{3}{|c|}{
Non-trivial commutation relations of $E_{8}$ according to} \\
\multicolumn{3}{|c|}{the decomposition given by eq.(\ref{84})}\\
\hline $\left[Q_{1},Q_{0}\right]=\sqrt{30}Q_{1}$ &
$\left[Q_{1},Q'_{0}\right]=\sqrt{10}Q_{1}$ &
$\left[Q_{2},Q_{0}\right]=-\sqrt{30}Q_{2}$ \\
$\left[Q_{2},Q'_{0}\right]=\sqrt{10}Q_{2}$ &
$\left[Q_{3},Q_{0}\right]=0$ &
$\left[Q_{3},Q'_{0}\right]=-2\sqrt{10}Q_{3}$ \\
$\left[Q_{1},Q^{1}\right]=-\sqrt{30}Q_{0}-\sqrt{10}Q'_{0}$ &
$\left[Q_{2},Q^{2}\right]=\sqrt{30}Q_{0}-\sqrt{10}Q'_{0}$ &
$\left[Q_{3},Q^{3}\right]=2\sqrt{10}Q'_{0}$ \\
$\left[Q_{1},Q_{2}\right]=\sqrt{20}Q^{3}$ &
$\left[Q_{2},Q_{3}\right]=\sqrt{20}Q^{1}$ &
$\left[Q_{3},Q_{1}\right]=\sqrt{20}Q^{2}$ \\
$\left[Q_{1i},Q_{0}\right]=\sqrt{30}Q_{1i}$ &
$\left[Q_{1i},Q'_{0}\right]=\sqrt{10}Q_{1i}$ &
$\left[Q_{2i},Q_{0}\right]=-\sqrt{30}Q_{2i}$ \\
$\left[Q_{2i},Q'_{0}\right]=\sqrt{10}Q_{2i}$ &
$\left[Q_{3i},Q_{0}\right]=0$ &
$\left[Q_{3i},Q'_{0}\right]=-2\sqrt{10}Q_{3i}$ \\
$\left[Q_{1i},Q_{2j}\right]=\sqrt{20}d_{ijk}Q^{3k}$ &
$\left[Q_{2i},Q_{3j}\right]=\sqrt{20}d_{ijk}Q^{1k}$ &
$\left[Q_{3i},Q_{1j}\right]=\sqrt{20}d_{ijk}Q^{2k}$ \\
$\left[Q^{\alpha},Q^{\beta}\right]=2ig^{\alpha\beta\gamma}Q^{\gamma}$
& $\left[Q^{\alpha},Q_{0}\right]=0$ &
$\left[Q^{\alpha},Q'_{0}\right]=0$ \\
$\left[Q^{\alpha},Q_{1i}\right]=-(G^{\alpha})^{j}_{i}Q_{1j}$ &
$\left[Q^{\alpha},Q_{2i}\right]=-(G^{\alpha})^{j}_{i}Q_{2j}$ &
$\left[Q^{\alpha},Q_{3i}\right]=-(G^{\alpha})^{j}_{i}Q_{3j}$\\\hline
\end{tabular} \\ \vspace{.2in}

\begin{tabular}{|c|}\hline
Table 3 \\ Further non-trivial commutation relations of $E_{8}$\\
according to the decomposition given in eq.(\ref{84})\\ \hline
$\left[Q_{1i},Q^{1j}\right]=-\frac{1}{6}(G^{\alpha})^{j}_{i}Q^{\alpha}
-\sqrt{30}\delta^{j}_{i}Q_{0}-\sqrt{10}\delta^{j}_{i}Q'_{0}$ \\
$\left[Q_{2i},Q^{2j}\right]=-\frac{1}{6}(G^{\alpha})^{j}_{i}Q^{\alpha}
+\sqrt{30}\delta^{j}_{i}Q_{0}-\sqrt{10}\delta^{j}_{i}Q'_{0}$ \\
$\left[Q_{3i},Q^{3j}\right]=-\frac{1}{6}(G^{\alpha})^{j}_{i}Q^{\alpha}
+2\sqrt{10}\delta^{j}_{i}Q'_{0}$\\ \hline
\end{tabular} \\ \vspace{.2in}

The normalization of the generators in tables 2, 3 is
$$Tr(Q_{0}Q_{0})=Tr(Q'_{0}Q'_{0})=Tr(Q_{1}Q^{1})=Tr(Q_{2}Q^{2})=Tr(Q_{3}Q^{3})=2.$$
$$Tr(Q_{1i}Q^{1j})=Tr(Q_{2i}Q^{2j})=Tr(Q_{3i}Q^{3j})=2\delta^{j}_{i}.$$
$$Tr(Q^{\alpha}Q^{\beta})=12\delta^{\alpha\beta}.$$

\section{A three generation non-supersymmetric example:\\ Reduction of the $G=E_{8}$ over $B=CP^{2} \times S^{2}$}
In the present section we would like to present an interesting three
family model in which the original higher-dimensional supersymmetry
disappears completely in the process of dimensional reduction. This
is the case when we reduce a SYM on symmetric coset spaces. Then the
organization of the ten-dimensional Lagrangian in terms of
four-dimensional superfields is not possible and the reduction must
be carried out in the components. The rules are stated separately
for bosons and fermions and the surviving four-dimensional fields
belong to different representations of the four-dimensional gauge
group $H$. Following refs~\cite{Forgacs:1979zs, Kapetanakis:hf,
Manton:1981es, Manousselis:2001re} in the next paragraphs we
summarize the results.

To find the four-dimensional gauge group we embed $R$ in $G$
\begin{eqnarray}
G &\supset& R_{G} \times H,\nonumber \\
H&=&C_{G}(R_{G}).
\end{eqnarray}
Then $H$, the centralizer of the image of $R$ in $G$, is the
four-dimensional gauge group.

The scalar fields that are obtained from the higher components of
the vector field are obtained as follows. First we embed  $R$ in $S$
and decompose the adjoint of $S$ under $R$
\begin{eqnarray} \label{dec11}
S &\supset& R \nonumber \\ adjS &=& adjR+\sum s_{i}.
\end{eqnarray}
Then we embed $R$ in $G$ and decompose the adjoint of $G$ under $R
\times H $,
\begin{eqnarray} \label{dec22}
G &\supset& R_{G} \times H \nonumber \\
 adjG &=&(adjR,1)+(1,adjH)\nonumber\\&&+\sum(r_{i},h_{i}).
\end{eqnarray}
The rule is that when $$s_{i}=r_{i},$$ i.e. when we have two
identical representation of $R$ in the decompositions (\ref{dec1}),
(\ref{dec2}) there is an $h_{i}$ multiplet of scalar fields that
survives in four dimensions.

To find the representation of $H$ under which the four-dimensional
fermions transform, we have to decompose the representation $F$ of
$G$ to which the fermions belong, under $R_{G} \times H$, i.e.
\begin{equation}\label{eq:F}
F= \sum (t_{i},h_{i}),
\end{equation}
and the spinor of $SO(d)$ under $R$
\begin{equation}\label{eq:d}
\sigma_{d} = \sum \sigma_{j}.
\end{equation}
Then for each pair $t_{i}$ and $\sigma_{i}$, where $t_{i}$ and
$\sigma_{i}$ are identical irreducible representations, survives
an $h_{i}$ multiplet of spinor fields in the four-dimensional
theory. In our case we have $F=adjG$.

Next let us discuss a specific model which appears to be of
particular interest. Specifically we choose a ten-dimensional SYM
theory based on $G=E_{8}$ which we reduce over the coset space
$S/R=CP^{2} \times S^{2} = (SU(3)/SU(2) \times U(1)) \times
(SU(2)/ U(1))$. The embedding of $R=SU(2) \times U(1) \times U(1)$
in $E_{8}$ is given by the decomposition
$$E_{8} \supset SU(2) \times U(1) \times U(1) \times SO(10)$$
\begin{eqnarray}\label{e8}
248=(1,45)_{(0,0)}+(3,1)_{(0,0)}+(1,1)_{(0,0)}+(1,1)_{(0,0)}+(1,1)_{(2,0)}
\nonumber \\
+(1,1)_{(-2,0)}+(2,1)_{(1,2)}+(2,1)_{(-1,2)}+(2,1)_{(-1,-2)}+(2,1)_{(1,-2)}
\nonumber \\
+(1,10)_{(0,2)}+(1,10)_{(0,-2)}+(2,10)_{(1,0)}+(2,10)_{(-1,0)}
\nonumber \\ +(2,16)_{(0,1)}+(1,16)_{(1,-1)}+(1,16)_{(-1,-1)}
\nonumber \\
+(2,\overline{16})_{(0,-1)}+(1,\overline{16})_{(-1,1)}+(1,\overline{16})_
{(1,1)},
\end{eqnarray}
where the $R=SU(2) \times U(1) \times U(1)$ is identified with the
one appearing in the following decomposition of maximal subgroups
$$ SO(6) \supset SU(2) \times SU(2) \times U(1) \supset SU(2)
\times U(1) \times U(1)$$ and the $SU(2) \times SU(2) \times U(1)$
in $E_{8}$ is the maximal subgroup of $SO(6)$ appearing in the
decomposition $$E_{8} \supset SO(6) \times SO(10) $$
\begin{equation}
248 = (15,1)+(1,45)+(6,10)+(4,16)+(\overline{4},\overline{16}).
\end{equation}
We find that the four-dimensional gauge group is
$H=C_{E_{8}}(SU(2) \times U(1) \times U(1))=SO(10) \times U(1)
\times U(1)$. The vector and spinor content under $R$ of the
specific coset can be found in table 7. Choosing $a=b=1$ we find
that the scalar fields of the four-dimensional theory transform as
$10_{(0,2)}$, $10_{(0,-2)}$, $10_{(1,0)}$, $10_{(-1,0)}$ under
$H$. Also, we find that the fermions of the four-dimensional
theory are the following left-handed multiplets of $H$:
$16_{(-1,-1)}$, $16_{(1,-1)}$, $16_{(0,1)}$.

To determine the potential we examine further the decomposition of
the adjoint of  $S=SU(3) \times SU(2)$ under $R= SU(2) \times U(1)
\times U(1)$ The specific coset has the direct product structure
$$ \frac{SU(3)}{SU(2) \times U(1)} \times \frac{SU(2)}{U(1)} $$
and the embeddings are $$SU(3) \supset SU(2) \times U(1) $$ and $$
SU(2) \supset U(1)$$ The relevant decompositions of the adjoint of
$SU(3)$ and $SU(2)$ are respectively
\begin{eqnarray}
8 &=& 1_{0} + 3_{0} + 2_{1} + 2_{-1} \nonumber \\ 3 &=& (0) + (2)
+ (-2)
\end{eqnarray}
The decomposition of the generators of $SU(3)$ and $SU(2)$ are
\begin{equation}\label{su3}
Q_{0}, Q_{\rho}, Q_{(+)a}, Q_{(-)a}
\end{equation}
and
\begin{equation}\label{su2}
Q'_{0}, Q_{(+)}, Q_{(-)}
\end{equation}
respectively. The commutation relations of $SU(2)$ are
\begin{equation}
\left[ Q_{(+)}, Q_{(-)} \right] = 2 Q'_{0},\ \ \left[ Q'_{0},
Q_{(\pm)} \right] = \pm 2 Q_{(\pm)}
\end{equation}
with the normalization $Tr(Q_{(+)}Q_{(-)}) = Tr(Q_{0}^{'2}) = 2 $,
while the commutation relations of $SU(3)$ are given in table 4.

\begin{centering}
\begin{tabular}{|l|l|}\hline
\multicolumn{2}{|c|}{Table 4}\\ \multicolumn{2}{|c|}{Non-trivial
commutation relations of $SU(3)$}\\ \hline $\left[
Q_{\rho},Q_{\sigma} \right] = 2i \epsilon_{\rho \sigma
\tau}Q_{\tau}$ & $\left[ Q_{0},Q_{(+)a} \right] = Q_{(+)a}$
\\
$\left[ Q_{\rho},Q_{(+)a} \right] = (\tau_{\rho})^{b}_{a}Q_{(+)b}$
& $\left[ Q_{(+)a},Q_{(-)b} \right] = (\tau_{\rho})_{b \ a}
Q_{\rho} + \delta_{a \ b}Q_{0}$
\\ \hline
\end{tabular} \\ \vspace{.2in}
\end{centering}

The normalization of the generators in table 4 is
$$Tr(Q_{\rho}Q_{\sigma})=2\delta^{\rho \ \sigma},
Tr(Q_{(+)a}Q_{(-)b})=2\delta_{a \ b}, Tr(Q_{0}^{2}) = 2 .
$$ According to the decompositions (\ref{su3}) and (\ref{su2}), we denote the
constrained scalar fields by
\begin{equation}\label{Confields}
\phi_{(+)}, \phi_{(-)}, \phi_{(+)a}, \phi_{(-)a}
\end{equation}
and then the constraints to be solved become
\begin{eqnarray}\label{constraints}
\left[ \phi'_{(0)}, \phi_{(\pm)}\right] &=& \pm 2 \phi_{(\pm)},
\nonumber\\ \left[ \phi_{(0)}, \phi_{(\pm)a} \right] &=& \pm
\phi_{( \pm )a }, \nonumber\\ \left[ \phi_{\rho}, \phi_{(\pm)a}
\right] &=& \pm (\tau_{\rho})_{a}^{b}\phi_{(\pm)b}.
\end{eqnarray}
We find that the potential of any ten-dimensional theory reduced
over $CP^{2} \times S^{2}$  written in terms of the fields
(\ref{Confields}) takes the form
\begin{eqnarray}\label{potential}
V &=&\left( \frac{4(\bar{\Lambda}^{2} + \Lambda^{2})}{R_{1}^{2}} +
\frac{4 \Lambda'^{2}}{R_{2}^{2}} \right) -
\frac{1}{2R_{1}^{4}}(\tau_{\rho})^{ab}Tr \phi_{\rho}[\phi_{(+)a},
\phi_{(-)b}] - \frac{1}{2R_{1}^{4}}\delta^{a}_{b}Tr
\phi_{(0)}[\phi_{(+)a}, \phi_{(-)b}]\nonumber \\ & & +
\frac{1}{4R_{1}^{4}} Tr \left( [\phi_{(+)a},
\phi_{(+)b}][\phi_{(-)a}, \phi_{(-)b}] + [\phi_{(-)a},
\phi_{(+)b}][\phi_{(+)a}, \phi_{(-)b}] \right) \nonumber\\&& -
\frac{2}{R_{2}^{4}}Tr \phi'_{(0)}[\phi_{(+)}, \phi_{(+)}] +
\frac{1}{2R_{2}^{4}} Tr \left([\phi_{(+)},  \phi_{(-)}]^{2} \right),
\end{eqnarray}
where $R_{1}$ and $R_{2}$ are the coset space radii corresponding to
the factors $CP^{2}$ and $S^{2}$ respectively.

To proceed in determining the four-dimensional potential of our
specific example with $G=E_{8}$, we use the embedding (\ref{e8}) of
$SU(2) \times U(1) \times U(1)$ in $E_{8}$ and divide its generators
accordingly,
\begin{eqnarray}\label{gen76}
Q_{E_{8}} &=& \{ G^{\alpha \beta},\ \  G^{\rho},\ \  G,\ \  G',\ \
G_{(+)},\ \  G_{(-)},\nonumber\\&& K_{(+)a},\ \  K_{(-)a},\ \
J_{(+)a},\ \  J_{(-)a}, \nonumber \\ &&Q_{(+)i},\ \  Q_{(-)i},\ \
Q_{(+)ai},\ \  Q_{(-)ai} \nonumber \\ &&F_{(+)m},\ \
\bar{F}_{(-)}^{m},\ \  H_{(+)m},\ \ \bar{H}_{(-)}^{m},\ \
F_{(+)am}, \bar{F}_{(-)a}^{m} \},
\end{eqnarray}
with the indices $\alpha \beta, i$ and $m$ belong to the $45, 10$
and $16$ of $SO(10)$, while $\rho$ and $a$ to the $3$ and $2$ of
$SU(2)$ respectively. The non-trivial commutation relations of the
generators (\ref{gen76}) are given in tables 4 and 5.  \\

\begin{tabular}{|l|l|l|}\hline
\multicolumn{3}{|c|}{Table 4}\\  \multicolumn{3}{|c|}{ Non-trivial
commutation relations of $E_{8}$ according }
\\ \multicolumn{3}{|c|}{the decomposition given by eq.
(\ref{e8})}\\ \hline $ $  & $\left[G^{\alpha \beta}, G\right] = 0$
& $\left[G^{\alpha \beta}, G'\right] = 0$ \\ $\left[G^{\alpha
\beta}, Q_{(+)i}\right]= (G^{\alpha \beta})_{i}^{j}Q_{(+)j}$ &
$\left[G^{\alpha \beta}, Q_{(+)ai}\right]= (G^{\alpha \beta
})_{i}^{j}Q_{(+)aj}$ & $\left[G^{\alpha \beta}, F_{(+)m} \right] =
(G^{\alpha \beta})_{m}^{n}F_{(+)n}$ \\ $\left[G^{\alpha \beta},
H_{(+)m}\right] = (G^{\alpha \beta})_{m}^{n}H_{(+)n}$ &
$\left[G^{\alpha \beta}, F_{(+)a m} \right] = (G^{\alpha \beta
})_{m}^{n}F_{(+)a n}$ & $ $ \\ $\left[G^{\rho}, G^{\sigma}\right]
=2i \epsilon^{\rho \sigma \tau} G^{\tau}$ & $\left[G^{\rho},
G\right] =0$ & $\left[G^{\rho}, G'\right]=0$ \\ $\left[G^{\rho},
K_{(+)a} \right] = (\tau^{\rho})_{a}^{b} K_{(+)b}$ &
$\left[G^{\rho}, J_{(+)a}\right] = (\tau^{\rho})_{a}^{b} J_{(+)b}$
& $\left[G^{\rho}, Q_{(+)ai}\right] = (\tau^{\rho})_{a}^{b}
Q_{(+)bi}$ \\ $\left[G^{\rho}, F_{(+)am}\right] =
(\tau^{\rho})_{a}^{b} F_{(+)bm}$ & $ $ & $ $ \\ $\left[Q_{(+)i},
Q_{(+)aj}\right] = \frac{1}{2}\delta_{ij}K_{(+)a}$ &
$\left[Q_{(+)i}, Q_{(-)aj}\right] =
\frac{1}{2}\delta_{ij}J_{(+)a}$ & $\left[Q_{(+)ai},
Q_{(+)bj}\right] = 0$ \\ $\left[Q_{(+)i},F_{(+)m}\right] =
C_{imn}\bar{H}_{(-)}^{n} $ & $ \left[Q_{(+)i}, H_{(+)m}\right] =
-C_{imn}\bar{F}_{(-)}^{n} $ & $\left[Q_{(+)i}, \bar{F}_{(-)b}^{m}
\right] = 0$ \\ $\left[Q_{(+)ai}, H_{(+)m} \right] =
C_{imn}\bar{F}_{(-)a}^{n}$ & $ \left[Q_{(+)ai}, F_{(+)bn}\right] =
-\delta_{ab} C_{imn}\bar{H}_{(-)}^{n}$ & $\left[Q_{(-)ai},
F_{(+)m}\right] = C_{imn} \bar{F}_{(-)a}^{n}$ \\
$\left[Q_{(-)ai},F_{(+)bm}\right] = -\delta_{ab} C_{imn}
\bar{F}_{(-)}^{n} $ & $ $ & $ $ \\ $ \left[F_{(+)m},
H_{(+)n}\right] = C^{i}_{mn}Q_{(-)i}$ & $\left[F_{(+)m},
F_{(+)an}\right] = C^{i}_{mn}Q_{(+)ai}$ & $ \left[H_{(+)m},
F_{(+)an}\right] = C^{i}_{mn}Q_{(-)ai}$ \\ $\left[F_{(+)am},
F_{(+)bn}\right] = 0$ & $ $ & $ $
\\ $\left[F_{(+)m}, \bar{H}_{(-)}^{n}\right] =
\delta_{m}^{n}G_{(+)}$ & $ \left[F_{(+)m}, \bar{F}_{(-)a}^{n}
\right] = 0$ & $\left[H_{(+)m}, \bar{F}_{(-)}^{n} \right] =
\delta_{m}^{n}G_{(-)}$ \\ $\left[H_{(+)m}, \bar{F}_{(-)a}^{n}
\right] = 0$ & $ \left[F_{(+)am}, \bar{F}_{(-)}^{n}\right] = 0$ &
$\left[F_{(+)am}, \bar{F}_{(-)b}^{n}\right] = 0$\\ $\left[G^{(')},
G_{(+)} \right] = 0$ & $\left[G^{(')}, K_{(+)a}\right] =
3K_{(+)a}$ & $\left[G^{(')}, J_{(+)a}\right] = 3J_{(+)a}$ \\
$\left[G^{(')}, Q_{(+)i}\right] = 2Q_{(+)i}$ & $\left[G^{(')},
Q_{(+)ai}\right] = Q_{(+)ai} $ & $\left[G^{(')}, F_{(+)m}\right] =
-F_{(+)m}$ \\ $\left[G^{(')}, H_{(+)m}\right] = -H_{(+)m}$ & $
\left[G^{(')}, F_{(+)am}\right] = -F_{(+)am} $ & $ $ \\ $
\left[G_{(+)}, G_{(-)}\right] =0$ & $\left[G_{(+)},
J_{(+)a}\right] = K_{(+)a}$ & $\left[G_{(+)}, K_{(-)a}\right] =
-J_{(-)a}$ \\ $\left[G_{(+)}, Q_{(-)ai}\right] = 0$ &
$\left[G_{(+)}, H_{(+)m}\right] = F_{(+)m}$ & $\left[G_{(+)},
\bar{F}_{(-)}^{m}\right] = -\bar{H}_{(-)}^{m}$ \\ $\left[K_{(+)a},
J_{(-)b}\right] = -\delta_{ab} G_{(+)}$ & $ $ & $ $ \\
$\left[K_{(+)a}, Q_{(-)i}\right]= \frac{1}{2} Q_{(+)ai}$ & $
\left[K_{(+)a}, Q_{(-)bj}\right] = \frac{1}{2} \delta_{ab}
Q_{(+)j}$ & $\left[J_{(+)a}, Q_{(-)j}\right] = \frac{1}{2}
Q_{(-)ai}$ \\ $ \left[J_{(+)a}, Q_{(+)bj}\right] = \frac{1}{2}
\delta_{ab} Q_{(+)j}$ &  $ $ & $ $ \\ $\left[K_{(+)a},
\bar{F}_{(-)b}^{m}\right] =0$ & $\left[K_{(+)a},
H_{(+)m}\right]=0$ & $\left[J_{(+)}, F_{(+)m} \right]=0$ \\
$\left[ J_{(+)a}, \bar{F}_{(-)b}^{m} \right]=0$ & $\left[G,
G'\right] =0$ & $ $ \\ \hline
\end{tabular} \\ \vspace{.2in}

\begin{tabular}{|c|}\hline
Table 5 \\ Further commutation relations of $E_{8}$\\ according to
the dcomposition given by eq.(\ref{e8})\\ \hline $\left[Q_{(+)i},
Q_{(-)j}\right] = \frac{1}{6}(G^{\alpha \beta })_{ij}G^{\alpha
\beta} + 2\delta_{ij}G + 2\delta_{ij} G' $\\ $ \left[ Q_{(+)ai},
Q_{(-)bj}\right] = \frac{1}{6}\delta_{ab}(G^{\alpha
\beta})_{ij}G^{\alpha \beta} +
\delta_{ij}(\tau^{\rho})_{ab}G^{\rho}+ \delta_{ij}G + \delta_{ij}
G'$\\ $\left[F_{(+)m},\bar{F}_{(-)}^{n}\right] =
\frac{1}{6}(G^{\alpha \beta})_{m}^{n}G^{\alpha \beta} -
\delta_{m}^{n}G - \delta_{m}^{n}G'$\\ $ \left[H_{(+)m},
\bar{H}_{(-)}^{n} \right] = \frac{1}{6}(G^{\alpha \beta
})_{m}^{n}G^{\alpha \beta}- \delta_{m}^{n}G - \delta_{m}^{n}G'$ \\
$\left[F_{(+)am},\bar{F}_{(-)b}^{n}\right] =
\frac{1}{6}\delta_{ab}(G^{\alpha \beta})_{m}^{n}G^{\alpha \beta}+
(\tau^{\rho})_{ab}\delta_{m}^{n}G^{\rho}- \delta_{ab}
\delta_{m}^{n}G - \delta_{ab} \delta_{m}^{n}G'$\\ $\left[K_{(+)a},
K_{(-)b}\right] = (\tau^{\rho})_{ab}G^{\rho}+ 3\delta_{ab}G +
3\delta_{ab} G'$\\ $ \left[J_{(+)a}, J_{(-)b}\right]
=(\tau^{\rho})_{ab}G^{\rho}+ 3\delta_{ab}G + 3\delta_{ab} G'$\\
$\left[G^{\alpha \gamma}, G^{\beta \delta} \right]= 2i(G^{\alpha
\delta} \delta^{\gamma \beta} - G^{\gamma \delta} \delta^{\alpha
\beta} + G^{\gamma \beta} \delta^{\alpha \delta}).$ \\ \hline
\end{tabular} \\ \vspace{.2in}

The normalization of the generators in tables 4 and 5 is $Tr
G^{\alpha\beta}G_{\gamma\delta}=
12\delta^{\alpha}_{\gamma}\delta^{\beta}_{\delta}$, $TrG^{\rho
\sigma} = 2 \delta^{\rho \sigma}$, $TrG_{(+)}G_{(-)} = TrG^{2} =
TrG'^{2} =2$, $TrK_{(+)a}K_{(-)b}= 2 \delta_{ab}$,
$TrQ_{(+)i}Q_{(-)j} = 2\delta_{ij}$, $TrQ_{(+)ai}Q_{(-)bj} = 2
\delta_{ab} \delta_{ij}$, $TrF_{(+)m}\overline{F}_{(-)}^{n}= 2
\delta_{m}^{n}$ $TrH_{(+)m}\overline{H}_{(-)}^{n}= 2
\delta_{m}^{n}$, $TrF_{(+)am}\overline{F}_{(-)b}^{n}= 2
\delta_{ab}
\delta_{m}^{n}$.\\

The solution of the constraints (\ref{constraints}) which will
provide us with the genuine four-dimensional scalar fields is given
by
$$\phi'_{(0)} = \Lambda' G', \phi_{(0)} = \Lambda G, \phi_{\rho} =
\bar{\Lambda} G_{\rho}, $$
\begin{equation}\label{genuine}
\phi_{(+)} = R_{1} \varphi^{i}Q_{(+)i}, \phi_{(-)} = R_{1}
\bar{\varphi}^{i}Q_{(-)i}, \phi_{(+)a} = \sigma^{i}Q_{(+)ai},
\phi_{(-)a} = \bar{\sigma}^{i}Q_{(-)ai},
\end{equation}
with $\Lambda = \Lambda' = \bar{\Lambda} =1$. Eventually the
potential (\ref{potential}) written in terms of the scalar fields
(\ref{genuine}) takes the form
\begin{eqnarray}\label{potential73}
V &=& \left(\frac{8}{R_{1}^{4}} + \frac{4}{R_{2}^{4}} \right) -
\frac{4}{R_{1}^{2}}(\sigma^{i} \bar{\sigma}_{i}) -
\frac{8}{R_{2}^{2}}(\varphi^{i} \bar{\varphi}_{i})\nonumber\\ &+&
\frac{19}{3}(\varphi^{i} \bar{\varphi}_{i})^{2} + \frac{16}{3}
(\sigma^{i} \bar{\sigma}_{i})^{2} \nonumber\\ &+&
\frac{1}{3}(\bar{\varphi}_{i}\bar{\varphi}_{i})(\varphi^{j}\varphi^{j})
+
\frac{1}{3}(\bar{\sigma}_{i}\bar{\sigma}_{i})(\sigma^{j}\sigma^{j}).
\end{eqnarray}

In order to examine further the present model we should employ again
the flux mechanism and study the minimization of the potential
(\ref{potential73}). To keep unchanged the number of families
background gauge configurations with non-trivial winding number are
required as in the example discussed in section 4.

\begin{centering}
\begin{tabular}{|l l l|} \hline
\multicolumn{3}{|c|}{Table 7}\\ \multicolumn{3}{|c|}{
Six-dimensional symmetric cosets with $rankS=rankR$}\\ \hline
$S/R$ & $SO(6)$ vector & $SO(6)$ spinor \\ \hline $SO(7)/SO(6)$ &
$6$ & $4$ \\ $SU(4)/SU(3) \times U(1)$ & $3_{-2}+\overline{3}_{2}$
& $1_{3}+3_{-1}$ \\  $Sp(4)/(SU(2) \times U(1))_{max}$ &
$3_{-2}+3_{2}$ & $1_{3}+3_{-1}$ \\ $SU(3) \times SU(2)/SU(2)
\times U(1) \times U(1)$ & $1_{0,2a}+1_{0,-2a}$ &
$1_{b,-a}+1_{-b,-a}$ \\  & $+2_{b,0}+2_{-b,0}$ & $+2_{0,a}$ \\
$Sp(4) \times SU(2)/SU(2) \times SU(2) \times U(1)$ &
$(1,1)_{2}+(1,1)_{-2}$ & $(2,1)_{1}+(1,2)_{-1}$ \\  & $+(2,2)_{0}$
& \\ $(SU(2)/U(1))^{3}$ & $(2a,0,0)+(-2a,0,0)$ &
$(a,b,c)+(-a,-b,c)$ \\  & $+(0,2b,0)+(0,-2b,0)$  &
$+(-a,b,-c)+(a,-b,-c)$ \\ & $(0,0,2c)+(0,0,-2c)$ & \\ \hline
\end{tabular} \\ \vspace{.2in}
\end{centering}
\begin{centering}
\begin{tabular}{|l l l|} \hline
\multicolumn{3}{|c|}{Table 8}\\ \multicolumn{3}{|c|}{
Six-dimensional non-symmetric cosets with $rankS=rankR$}\\ \hline
$S/R$ & $SO(6)$ vector & $SO(6)$ spinor \\ \hline $G_{2}/SU(3)$ &
$3+\overline{3}$ & $1+3$ \\ $Sp(4)/(SU(2) \times U(1))_{non-max}$
& $1_{2}+1_{-2}+2_{1}+2_{-1}$ & $1_{0}+1_{2}+2_{-1}$ \\
$SU(3)/U(1) \times U(1)$ & $(a,c)+(b,d)+(a+b,c+d)$ &
$(0,0)+(a,c)+(b,d)$ \\ & $+(-a,-c)+(-b,-d)$ & $+(-a-b,-c-d)$ \\ &
$+(-a-b,-c-d)$ &  \\ \hline
\end{tabular} \\ \vspace{.2in}
\end{centering}

\section{Conclusions}
In the present work we have presented the details of the dimensional
reduction of  general ten-dimensional ${\cal N} =1$ supersymmetric
theories over coset spaces using the superfield formulation of the
theory.

In this way we have extended and generalized our previous results
concerning the supersymmetry breaking of ten-dimensional theories
in the process of dimensional reduction found using the components
formalism and a case by case study. Specifically we have shown
that ${\cal N}=1$, ten dimensional supersymmetric YM theories lead
in four dimensions, after reduction over coset spaces, either to a
softly broken ${\cal N}=1$ supersymmetric theory if the cosets
used are non-symmetric or to a non-supersymmetric theory if the
cosets used are symmetric.

We have also presented two very interesting three family models
resulting from a ${\cal N}=1$ supersymmetric $E_{8}$ gauge theory in
ten dimensions. The first one was reduced over the non-symmetric
coset space $SU(3)/U(1) \times U(1)$ and led in four dimensions to a
softly broken ${\cal N}=1$ gauge theory based on $E_{6}$-type gauge
group. This model has been examined earlier using the components
formalism and was reexamined here using superfields. The second
model concerns the reduction of the same ${\cal N}=1$,
ten-dimensional $E_{8}$ gauge theory over the symmetric coset space
$CP^{2} \times S^{2}$. The resulting four-dimensional gauge theory
is non-supersymmetric and is based on a $SO(10)$-type gauge group.
The details of this dimensional reduction are presented for the
first time.

Finally we have pointed out how the above very interesting three
family models at the GUT scale can be broken further using the
Wilson flux mechanism, while keeping the multiplicity of chiral
fermions. We plan to discuss the details of the phenomenological
analysis of both models in a future publication.

\section*{Acknowledgements}
We would like to thank A.~Sagnotti and A.~Pomarol for discussions
and in particular P.~Forgacs for extensive discussions and reading
the manuscript. The work of PM is supported by a
``$\Pi$Y$\Theta$A$\Gamma$OPA$\Sigma$" postdoctoral fellowship and
partially by the NTUA programme for fundamental research
``$\Theta$A$\Lambda$H$\Sigma$". The work of GZ is partially
supported by the programmes of Ministry of Education
``$\Pi$Y$\Theta$A$\Gamma$OPA$\Sigma$ " and
``HPAK$\Lambda$EITO$\Sigma$ " and by the NTUA programme
``$\Theta$A$\Lambda$H$\Sigma$".

\end{document}